\documentclass[letterpaper, 10 pt, conference]{ieeeconf}\IEEEoverridecommandlockouts
\usepackage{amsmath,amssymb,amsfonts}
\usepackage{algorithmic}
\usepackage{graphicx}
\usepackage{textcomp}
\usepackage{xcolor}
\usepackage{svg}
\usepackage{tabularx}
\usepackage{cuted}
\usepackage{multirow} 
\usepackage[font=small,labelfont=bf,labelsep=period]{caption}
\usepackage{xurl}
\usepackage{hyperref}
\hypersetup{
    colorlinks=true,
    linkcolor=black,
    filecolor=black,      
    urlcolor=blue,
    pdftitle={Overleaf Example},
    pdfpagemode=FullScreen,
    }
\def\mline{\vrule width4pt height2.5pt depth -2pt}
\def\dashed{\mline\hskip3.5pt\mline\thinspace}
\def\bdot{\raise.2em\hbox to .15em{.}}

\def\dotted{\bdot\, \bdot\,\bdot \thinspace}

\newtheorem{thm}{Theorem} 
\makeatletter

\renewcommand{\baselinestretch}{1} 

\def\BibTeX{{\rm B\kern-.05em{\sc i\kern-.025em b}\kern-.08em
    T\kern-.1667em\lower.7ex\hbox{E}\kern-.125emX}}

\def\rev#1{\textcolor{black}{#1}}

\title{\LARGE \bf
Stability Analysis of Thermohaline Convection With a Time-Varying Shear Flow Using the Lyapunov Method
}

\author{Kalin Kochnev$^{1}$ and Chang Liu$^{2}$
\thanks{$^{1}$School of Electrical and Computer Engineering, University of Connecticut, Storrs, CT 06269, USA
        {\tt\small kalinkochnev@proton.me}}
\thanks{$^{2}$School of Mechanical, Aerospace, and Manufacturing Engineering, University of Connecticut, Storrs, CT 06269, USA
        {\tt\small chang\_liu@uconn.edu}}%
\thanks{This research was supported by the University of Connecticut (UConn) Research Excellence Program (REP). K.K. acknowledges the support from the UConn Summer Undergraduate Research Fund (SURF) Awards. The computational work for this project was conducted using resources provided by the Storrs High-Performance Computing (HPC) cluster. We extend our gratitude to the UConn Storrs HPC and its team for their resources and support. The data and code of this paper are publicly available at \href{https://doi.org/10.5281/zenodo.17227640}{https://doi.org/10.5281/zenodo.17227640}. 
}
}
\begin{document}

\maketitle 
\begin{abstract}

This work demonstrates that the Lyapunov method can effectively identify the growth rate of a linear time-periodic system describing cold fresh water on top of hot salty water with a periodically time-varying background shear flow. We employ a time-dependent weighting matrix to construct a Lyapunov function candidate, and the resulting linear matrix inequalities are discretized in time using the forward Euler method. As the number of temporal discretization points increases, the growth rate predicted from the Lyapunov method or the Floquet theory will converge to the same value as that obtained from numerical simulations. Additionally, the Lyapunov method is used to analyze the most dangerous disturbance, and we also compare computational resource usage for the Lyapunov method, numerical simulations, and the Floquet theory. 

\end{abstract}
\allowdisplaybreaks

\section{Introduction}

Time dependence of flow speed is widely observed in the oceanic environment, such as the internal gravity waves and inertial waves \cite{alford2016near,sutherland2010internal}. Tidal currents are another source of time dependence, which are present in nearly all sub-ice-shelf observations and can significantly influence the ice melting rate \cite{washam2020tidal}. The time dependence of shear flow can also induce instabilities within parameter regimes where a steady shear flow will be stable \cite{radko2019thermohaline,radko2019instabilities}. For example, time-varying shear flows can interact with thermohaline convection (driven by the diffusivity difference between temperature and salinity) to induce thermohaline-shear instability \cite{radko2019thermohaline}. This thermohaline-shear instability was shown to exist in most high-latitude ocean regions, which plays an important role in mixing and staircase formation \cite{radko2019thermohaline}.

Stability analysis can provide key insights into pattern formation and the transition to chaotic and turbulent behaviors \cite{graham2021exact}, but the time dependence of background flow introduces additional challenges in stability analysis \cite{davis1976stability}. Quasi-steady assumption and eigenvalue analysis of the time-frozen base state are known to oversimplify the dynamics of a time-varying system, and thereby may miss instabilities induced by the time-dependence \cite{khalil2002nonlinear,knobloch2014stability}. Floquet theory \cite{floquet1883equations} is a powerful tool to analyze the stability of a periodic orbit, \rev{but Floquet methods are not applicable for nonlinear, non-periodic systems \cite{davis1976stability}.} Stability of time-varying systems can also be analyzed by numerical simulations \cite{radko2019thermohaline,radko2019instabilities}. However, simulations only support stability analysis through realizations of certain initial conditions, requiring exhaustive random search for nonlinear systems. For example, extensive simulations with more than 100,000 random initial conditions were used to characterize the region of attraction of a nine-mode shear flow model \cite{joglekar2015geometry}.

\rev{Unlike numerical simulation, Lyapunov-based stability analysis does not require extensive simulations, but directly characterizes the solution trajectory for a set of initial conditions.} The Lyapunov method has also been generalized to investigate the stability of time-varying (non-autonomous) systems \cite{khalil2002nonlinear,Boyd1994}. For example, a periodic Lyapunov function is introduced to analyze a discrete-time nonlinear periodically time-varying system \cite{bohm2012stability}, and stabilization of continuous-time periodic linear systems can be achieved by solving a periodic Lyapunov differential equation \cite{zhou2011periodic}. The Lyapunov method can also be applied to analyze nonlinear stability \cite{liu2020input}, nonlinear input-output properties \cite{wei2025nonlinear}, and non-periodically time-varying systems \cite{wei2025upper}, which can overcome limitations of Floquet theory. \rev{Here, we will analyze the special case of a linear \rev{periodically} time-varying system describing oceanic flows using the Lyapunov method, with the potential to extend to more general systems in mind. We demonstrate that the Lyapunov method can characterize the growth rate consistent with Floquet theory and numerical simulations for the defined system.}

\rev{Our model} describes thermohaline-shear instability \cite{radko2019thermohaline} where cold fresh water \rev{rests} on top of hot salty water \rev{that} interacts with a periodically time-varying shear flow. We will use a periodic weighting matrix $\boldsymbol{P}(t)$ to construct a Lyapunov function candidate $V(\boldsymbol{\psi},t)=\boldsymbol{\psi}^*\boldsymbol{P}(t)\boldsymbol{\psi}$, where $\boldsymbol{\psi}$ is the state variable, and the superscript $(\cdot)^*$ means the Hermitian of the argument. We formulate linear matrix inequalities (LMI) to obtain an upper bound of the growth rate, and the LMI is discretized in time by a forward Euler method. The resulting LMI will be coupled over time to solve the time-varying weighting matrix $\boldsymbol{P}(t)$ over one period. As the number of temporal discretization points increases, the growth rate predicted from the Lyapunov method or the Floquet theory converges to that obtained from numerical simulations. We also perform an eigendecomposition of the instantaneous weighting matrix $\boldsymbol{P}(t)$ to uncover the most dangerous disturbance of thermohaline-shear instability. Computational resource usage is compared among the Lyapunov method, numerical simulations, and Floquet theory. 

The paper is organized as follows. Sec. \ref{sec:problem_formulation} introduces the linear time-varying system that describes thermohaline-shear instabilities \cite{radko2019thermohaline}. Sec. \ref{sec:methods} presents how to compute growth rate using the Lyapunov method, numerical simulations, and the Floquet theory. Sec. \ref{sec:results} presents the results, and Sec. \ref{sec:conclusion} concludes this paper with a discussion of future directions.

\section{Problem Formulation}
\label{sec:problem_formulation}

\begin{figure}
    \centering
    \includegraphics[width=0.48\textwidth]{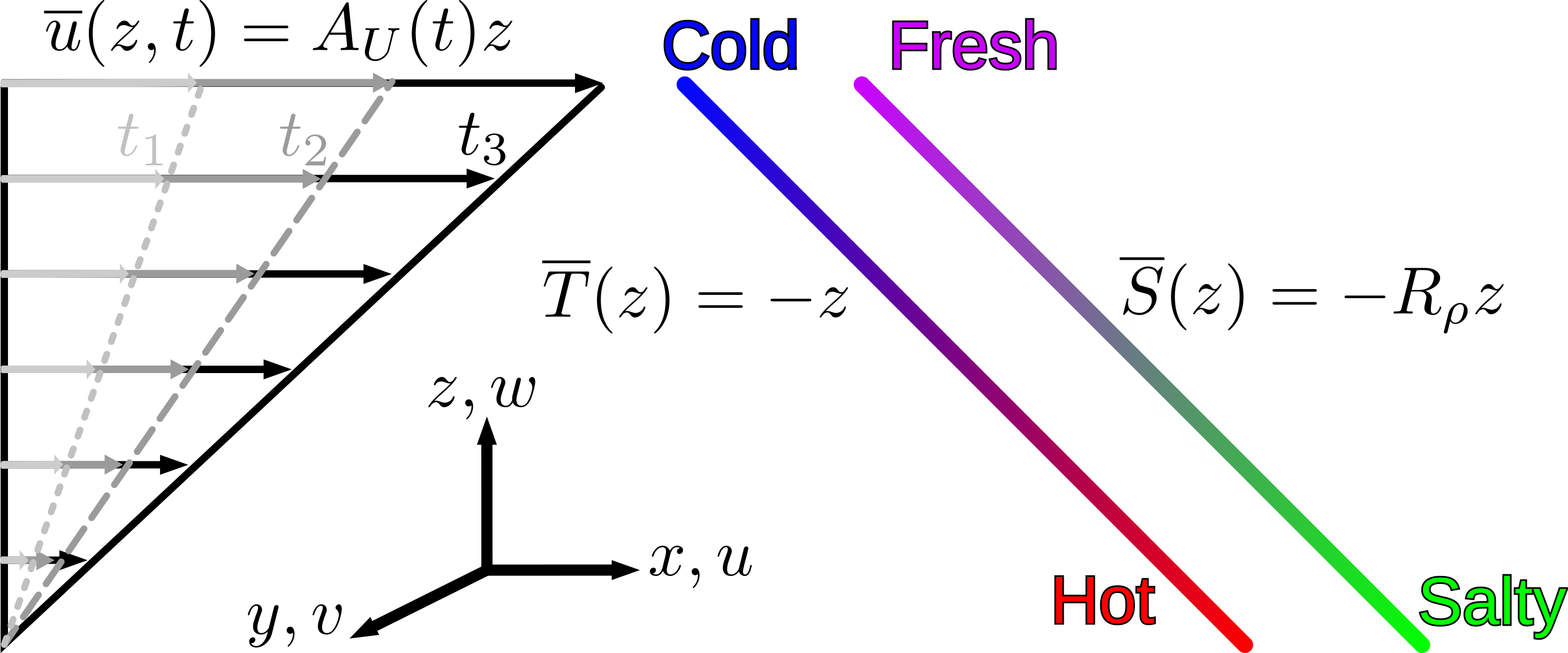}
    \caption{The linear time-varying model analyzed here describes cold fresh water on top of hot salty water with a time-varying shear flow $\overline{u}(z,t)$. The background shear $\overline{u}(z,t)=A_U(t)z$, temperature $\overline{T}(z)=-z$, and salinity $\overline{S}(z)=-R_\rho z$ vary linearly over $z$.}
    \label{fig:shearflow}
    \vspace{-3ex}
\end{figure}

We consider the flow configuration in Fig. \ref{fig:shearflow}, where we have a time-varying background shear flow $\overline{\boldsymbol{U}}_*(z_*,t_*)=[A_{U*}(t_*)z_*, A_{V*}(t_*)z_*, 0]^{\text{T}}$. We consider cold fresh water on top of hot salty water with a large-scale background temperature gradient $\partial_{z_*}\overline{T}_*<0$ and salinity gradient $\partial_{z_*}\overline{S}_*<0$. We assume that the density is influenced by temperature and salinity linearly $(\rho_* - \rho_{0*})/\rho_{0*}=\beta(S_*-S_{0*}) - \alpha(T_*-T_{0*})$, where $\alpha$ and $\beta$ are expansion and contraction coefficients. Here, $\rho_{0*}$, $T_{0*}$, and $S_{0*}$ are reference density, temperature, and salinity, respectively. We define dimensionless length, time, velocity, and pressure as $z=z_*/d$, $t=t_*/(d^2/\kappa_T)$, $u=u_*/(\kappa_T/d)$, and $p=p_*/(\rho_{0*}\nu \kappa_T/d^2)$ \cite{radko2019thermohaline}, where $d:=\left[\kappa_T \nu/(g\alpha |\partial_{z_*}\overline{T}_*|)\right]^{1/4}$, $\kappa_T$ is the thermal diffusivity, and $\nu$ is the viscosity. We normalize temperature and salinity by $T=T_*/(d\, |\partial_{z_*}\overline{T}_*|)$ and $S=\beta S_*/(\alpha\, d\, |\partial_{z_*}\overline{T}_*|)$. Then, we have dimensionless background temperature as $\overline{T}=-z$ and background salinity as $\overline{S}=-R_\rho z$ to model the transition from hot salty water on the bottom to cold fresh water on the top (Fig. \ref{fig:shearflow}). After linearization around the base state $(\overline{T},\overline{S},\overline{\boldsymbol{U}})$, we have the non-dimensional governing equation of fluctuations $(T,S,\boldsymbol{u})$ \cite{radko2019thermohaline} as:
\begin{subequations}
\label{eq:linearized_Navier_Stokes}
    \begin{align}
    \partial_t T+\overline{\boldsymbol{U}}{\cdot} \boldsymbol{\nabla}T-w&=\nabla^2 T,\\
    \partial_t S+\overline{\boldsymbol{U}}{\cdot} \boldsymbol{\nabla}S-R_\rho w&=\tau \nabla^2 S,\\
    \partial_t\boldsymbol{u}+\overline{\boldsymbol{U}}{\cdot} \boldsymbol{\nabla}\boldsymbol{u} + \boldsymbol{u}{\cdot}\boldsymbol{\nabla} \overline{\boldsymbol{U}}&\nonumber\\
    =\Pr[-\boldsymbol{\nabla}p& + \nabla^2 \boldsymbol{u}+(T-S)\boldsymbol{e}_z],\\
    \boldsymbol{\nabla}{\cdot} \boldsymbol{u}&=0. 
\end{align}
\end{subequations}
Here, we have the density ratio $R_\rho:=\beta \partial_{z_*}\overline{S}_*/(\alpha \partial_{z_*}\overline{T}_*)$, the Prandtl number $Pr:=\nu/\kappa_T$, and the diffusivity ratio $\tau:=\kappa_S/\kappa_T$, where $\nu$ is the viscosity and $\kappa_S$ is the salinity diffusivity. The unit vector in the vertical $z$ direction is denoted as $\boldsymbol{e}_z$. The dimensionless background shear flow is $\overline{\boldsymbol{U}}(z,t)=[\overline{u}(z,t),\overline{v}(z,t),0]^{\text{T}}=[A_U(t)z,A_V(t)z,0]^{\text{T}}$ with an unbounded vertical direction $z$, \rev{allowing us to consider the spatial domain $x,y,z\in (-\infty,\infty)$ with periodic boundary conditions for fluctuations $(T,S,\boldsymbol{u})$ in $x$, $y$, and $z$ directions.} Thus, we can perform the following solution ansatz \cite{radko2019thermohaline}:
\begin{align}
    (T,S,u,v,w) = \Re\left\lbrack \left(\hat{T}, \hat{S},\hat{u},\hat{v},\hat{w} \right)e^{\text{i}k {x} + \text{i}l {y} + \text{i}m {z}} \right\rbrack,
    \label{eq:fourier_ansatz}
\end{align}
where $\Re[\cdot]$ means the real part of the argument, $k$ and $l$ are respectively horizontal wavenumbers in $x$ and $y$ directions, and $m(t)$ is a time-varying vertical wavenumber:
\begin{align}
    m(t)= m_{0} - B_{U}(t)k - B_{V}(t)l. 
    \label{eq:m_t}
\end{align}
Here, $m_0$ is an initial vertical wavenumber, while $B_U(t)$ and $B_V(t)$ in \eqref{eq:m_t} are defined as
\begin{align}
    B_{U}(t) = \int_{0}^{t}A_{U}(t^\prime)dt^\prime,\;\;
    B_{V}(t) = \int_{0}^{t}A_{V}(t^\prime)dt^\prime. 
\end{align}

Using \eqref{eq:fourier_ansatz}, we can convert the linearized governing equation in \eqref{eq:linearized_Navier_Stokes} as a linear time-varying system expressed in \eqref{eq:ode}: 
\begin{equation}
    \dot{\boldsymbol{\psi}} = \boldsymbol{A}(t)\boldsymbol{\psi}, 
\label{eq:ode}
\end{equation}
where the state variable $\boldsymbol{\psi}=\begin{bmatrix}
    \hat{T} &
    \hat{S} &
    \hat{u} &
    \hat{v} &
    \hat{w}
\end{bmatrix}^\text{T}\in \mathbb{C}^{5\times 1}$ are the Fourier coefficients. \rev{The matrix $\boldsymbol{A}(t) \in \mathbb{R}^{5 \times 5}$} is defined in \eqref{eq:a_matrix} with $c(t)=k^2 + l^2 + m(t)^2$:
\begin{strip}
\noindent\rule{\textwidth}{0.4pt} 
\begin{equation}
\label{eq:a_matrix}
    \boldsymbol{A}(t) = \begin{bmatrix}
    - c(t) & 0 & 0 & 0 & 1 \\
    0 & - \tau c(t) & 0 & 0 & {R}_{\rho} \\
    \frac{- \Pr km(t)}{c(t)} & \frac{\Pr km(t)}{c(t)} &  - \Pr c(t) & 0 & \frac{2k[k A_{U}(t) + lA_{V}(t)]}{c(t)} - A_{U}(t) \\
    \frac{- \Pr lm(t)}{c(t)} & \frac{\Pr lm(t)}{c(t)} & 0 & - \Pr c(t)  & \frac{2l[kA_{U}(t) + lA_{V}(t)]}{c(t)} - A_{V}(t) \\
    - \frac{{\Pr m(t)}^{2}}{c(t)} + \Pr & \frac{{\Pr m(t)}^{2}}{c(t)} - \Pr & 0 & 0 & \frac{2m(t)[kA_{U}(t) + lA_{V}(t)]}{c(t)} - \Pr c(t) \\
\end{bmatrix}. 
\end{equation}
\noindent\rule{\textwidth}{0.4pt} 
\end{strip}
\noindent \rev{The detailed derivation of Eq. \eqref{eq:a_matrix} is available at Ref. \cite{kochnev2025thesis}.}

Here, we simplify this model as a two-dimensional monochromatic shear model by setting $l=A_V(t)=0$ and $A_U(t) = A_u \cos(\omega t)$ with the shear amplitude as $A_u = \sqrt{\frac{2 Pr (R_\rho -1)}{Ri}}$ \cite{radko2019thermohaline}. All dimensionless parameters used in this paper are summarized in Table \ref{table:sim_params}.

\begin{table}[h]
    \centering
    \begin{tabularx}{\columnwidth}{|l|l|X|}
    \hline
        \textbf{Variable} & \textbf{Value/Range} & \textbf{Description}\\
    \hline
        $Pr$     & 10                & Prandtl number\\
        $R_p$    & 2.0  		     & Density ratio \\
        $\tau$ 	 & 0.01 		     & Diffusivity ratio\\
        $Ri$    & 2.0  		     & Richardson number \\
        $\omega$ & 0.5			     & Oscillation frequency of background shear \\
        $k$		 & $[-0.5, 0.5]$ & Wavenumber in the $x$ direction\\
        $l$ 	 & 0			 &  Wavenumber in the $y$ direction\\
        $m_0$    & $[0, 1.5]$    & Initial wavenumber in the $z$ direction \\
        $n$ &     $[400,2000]$       & Temporal grid points of $\boldsymbol{A}(t)$ and $\boldsymbol{P}(t)$ \\
        $T$		 &    $4\pi$  & Oscillation period $T=2\pi/\omega$\\ 
        $n_\text{samples}$& 126           & Sample numbers for $k$ and $m_0$\\
        
    \hline
    \end{tabularx}
    \caption{A list of the parameters used in this work.}
    \vspace{-3ex}
    \label{table:sim_params}
\end{table}

\par\noindent

\section{Methods}
\label{sec:methods}
In this section, we present our formulation to compute the growth rate of the linear time-varying system in \eqref{eq:ode} using the Lyapunov method. We then show how to obtain the growth rate using numerical simulations and the Floquet theory, which will be used as a basis for comparison with the Lyapunov method. 
    
\subsection{Lyapunov stability analysis}
\label{subsec:lyapunov}

We will have the following Theorem \ref{thm:lmi_linear} to predict an upper bound on the growth rate. Here, $(\cdot)\preceq 0$ is negative semi-definiteness, $(\cdot )\succeq 0$ is positive semi-definiteness of the argument, and the superscript $(\cdot)^*$ means the Hermitian of the argument. 

\begin{thm}
Given a linear time-varying system $\dot{\boldsymbol{\psi}}=\boldsymbol{A}(t)\boldsymbol{\psi}$ \rev{with $\boldsymbol{\psi} \in \mathbb{C}^{5 \times 1}$. If} we can find a continuously differentiable Hermitian matrix $\boldsymbol{P}(t)\in \mathbb{C}^{5\times 5}$ and $\overline{\lambda}$ by
\begin{subequations}
\label{eq:lmi_all}
    \begin{align}
&\text{min  }\;\; \overline{\lambda} \\
&\text{subject to }\;\; \boldsymbol{P}(t) \succeq \epsilon \boldsymbol{I},\;\;\epsilon> 0,\;\;\forall t \in \mathbb{R},\label{eq:lmi_P_positive_difinite}\\
& \;\;\;\;\dot{\boldsymbol{P}}(t) + \boldsymbol{A}(t)\rev{^*}\boldsymbol{P}(t)+\boldsymbol{P}(t)\boldsymbol{A}(t) - 2 \overline{\lambda} \boldsymbol{P}(t) \preceq  0,
    \label{eq:lmi_linear_lambda}
\end{align}
\end{subequations}
then $\|\boldsymbol{\psi}(t)\|_2\leq C e^{\overline{\lambda} (t-t_0)}\|\boldsymbol{\psi}(t_0)\|_2$, where $C$ is a constant.  \label{thm:lmi_linear}
\end{thm}

\begin{proof}
Consider a function $V(\boldsymbol{\psi}, t) = \boldsymbol{\psi}^* \boldsymbol{P} (t) \boldsymbol{\psi}$ and compute its time derivative $\dot{V}$ as:
\begin{subequations}
\label{eq:proof_V_ineq_all}
    \begin{align}
        \dot{V} &= \boldsymbol{\psi}^*\dot{\boldsymbol{P}}(t)\boldsymbol{\psi}+\dot{\boldsymbol{\psi}}^*\boldsymbol{P}(t)\boldsymbol{\psi} + \boldsymbol{\psi}^*\boldsymbol{P}(t)\dot{\boldsymbol{\psi}} \\
        &= \boldsymbol{\psi}^*[ \dot{\boldsymbol{P}}(t)+ \boldsymbol{A}(t)^* \boldsymbol{P}(t) + \boldsymbol{P}(t) \boldsymbol{A}(t)] \boldsymbol{\psi}\\
        &\leq 2\overline{\lambda}\, \boldsymbol{\psi}^*\boldsymbol{P}(t)\boldsymbol{\psi} =2\overline{\lambda}\, V. \label{eq:proof_V_ineq}
\end{align}
\end{subequations}
Thus, we have $V(\boldsymbol{\psi}(t),t)\leq e^{2\overline{\lambda} (t-t_0)}V(\boldsymbol{\psi}(t_0),t_0)$. We then use the Rayleigh-Ritz theorem \cite[Theorem 4.2.2]{horn2012matrix} to obtain
\begin{align}
    \mu_\text{min}[\boldsymbol{P}(t)]  \boldsymbol{\psi}^* \boldsymbol{\psi}  &\leq V(\boldsymbol{\psi}, t) \leq   
    \mu_\text{max}[\boldsymbol{P}(t)] \boldsymbol{\psi}^* \boldsymbol{\psi},
    \label{eq:Rayleigh_Ritz}
\end{align}
where $\mu_{\text{min}}[\cdot]$ and $\mu_{\text{max}}[\cdot]$ represent the minimum and maximum eigenvalues of the argument. 

Combining the bound on $V(\boldsymbol{\psi},t)$ and Eq. \eqref{eq:Rayleigh_Ritz}, we obtain
     \begin{align}
         &\mu_\text{min}[\boldsymbol{P}(t)] \| \boldsymbol{\psi}(t) \| _2^2\leq  V(\boldsymbol{\psi}(t), t)\leq e^{2 \overline{\lambda} (t-t_0)} V(\boldsymbol{\psi}(t_0), t_0)\nonumber \\
         &
         \leq e^{2 \overline{\lambda} (t-t_0)} \mu_\text{max}[\boldsymbol{P}(t_0)] \|\boldsymbol{\psi}(t_0)\|_2^2.
     \end{align}
Thus, we have $ \|  \boldsymbol{\psi}(t) \|_2 
\leq   \sqrt{\frac{ \mu_\text{max}[\boldsymbol{P}(t_0)]}{\mu_\text{min}[\boldsymbol{P}(t)] }}  e^{\overline{\lambda} (t-t_0)} \|\boldsymbol{\psi}(t_0)\|_2 \leq \sqrt{\frac{ \mu_\text{max}[\boldsymbol{P}(t_0)]}{\epsilon }}  e^{\overline{\lambda} (t-t_0)} \|\boldsymbol{\psi}(t_0)\|_2$ using Eq. \eqref{eq:lmi_P_positive_difinite}. 
\end{proof}

\rev{Theorem \ref{thm:lmi_linear} will prove the stability if Eqs. \eqref{eq:lmi_P_positive_difinite} and \eqref{eq:lmi_linear_lambda} are feasible with $\overline{\lambda}=0$. It also allows us to analyze the growth rate for instabilities ($\overline{\lambda}>0$), which will be a main focus of this paper.} We then discretize $\boldsymbol{P}(t)$ as $\boldsymbol{P}(t_i)$ and discretize $\boldsymbol{A}(t)$ as $\boldsymbol{A}(t_i)$, $i=0,1,...,n$ with a fixed time step $\Delta t = t_{i+1} - t_i$. \rev{Because the time derivative cannot be directly implemented in our LMI solver,} we use the forward Euler method to approximate it as $\dot{\boldsymbol{P}} (t_i)=\frac{\boldsymbol{P}(t_{i+1}) - \boldsymbol{P}(t_i)}{\Delta t}$, leading to the discretized LMI as: 
\begin{subequations}
    \label{eq:lmi_all_discretized}
    \begin{align}
        &\text{min } \;\;\overline{\lambda} \\
        &\text{subject to }\;\; \boldsymbol{P}(t_i) \succeq\epsilon \boldsymbol{I}, \,\, \epsilon >0, \,\, \forall i = 0,1, ..., n\!-\!1, \\
        &\dot{\boldsymbol{P}}(t_i) + \boldsymbol{A}(t_i)^*\boldsymbol{P}(t_i)+\boldsymbol{P}(t_i)\boldsymbol{A}(t_i)-2\overline{\lambda}\boldsymbol{P}(t_i) \preceq 0. \label{eq:lmi_linear_lambda_discretized}
    \end{align}
\end{subequations}
For a time-periodic system, we can enforce periodicity on $\boldsymbol{P}(t)$ such that $\boldsymbol{P}(t_n)=\boldsymbol{P}(t_0)$ \cite{bohm2012stability,zhou2011periodic}, and we only need to consider one period for the discretized LMI. Thus, we select $t_0=0$ and $t_n=T$, where $T=\frac{2\pi}{\omega}$ is the oscillation period. \rev{Although it is possible to first discretize the state-space model in Eq. \eqref{eq:ode} and then search for a Lyapunov function for the discrete-time system, here we formulate an LMI in Theorem \ref{thm:lmi_linear} for a continuous-time system and then discretize the LMI, allowing us to define the upper bound of growth rate $\overline{\lambda}$ for a direct comparison with numerical simulations \cite{radko2019thermohaline} and Floquet theory. Moreover, as this work considers a linear system in \eqref{eq:ode}, the region of attraction will be the entire $\boldsymbol{\psi}\in \mathbb{C}^{5\times 1}$ for stable parameter regimes. }

\rev{In Theorem \ref{thm:lmi_linear}, the $\boldsymbol{P}(t)\in \mathbb{C}^{5\times 5}$ matrix is complex because the state variables $\boldsymbol{\psi}\in \mathbb{C}^{5\times 1}$ are Fourier coefficients and in general complex. However, because $\boldsymbol{A}(t)\in \mathbb{R}^{5\times 5}$ here in \eqref{eq:a_matrix} is a real matrix (which is not always the case for fluid systems, e.g., \cite{wei2025upper}), we can assume the weighting matrix $\boldsymbol{P}(t)\in \mathbb{R}^{5\times 5}$ and $\boldsymbol{P}(t_i)\in \mathbb{R}^{5\times 5}$ as real symmetric matrices without loss of generality.} The value of $\epsilon$ does not influence the upper bound of the growth rate $\overline{\lambda}$ due to the homogeneity of the constraints in \eqref{eq:lmi_linear_lambda} and \eqref{eq:lmi_linear_lambda_discretized}, and thus, we set $\epsilon=1$ without loss of generality. We use YALMIP \cite{Lofberg2004} to formulate our LMI in \eqref{eq:lmi_all_discretized} and use MOSEK \cite{Mosek2024} as the semi-definite programming (SDP) solver. We first check the stability of the system by checking the feasibility of constraints in \eqref{eq:lmi_all_discretized} with $\overline{\lambda}=0$, which is much quicker than calculating the value of $\overline{\lambda}$. If infeasible, we then solve \eqref{eq:lmi_all_discretized} to minimize the upper bound of the growth rate $\overline{\lambda}>0$ through bisection search over $\overline{\lambda}$. \rev{All default settings are used for the YALMIP solver, where the bisection absolute gap tolerance of $10^{-5}$ is used.} 

\subsection{Numerical simulation} \label{subsec:dns}
The linear time-varying system in \eqref{eq:ode} was numerically integrated with \texttt{ode45} command in MATLAB for 50 periods, and we sample results $\boldsymbol{\psi}(t)$ at a time step $\Delta t=\frac{50 T}{2000 }$. The non-dimensional perturbation energy in \eqref{eq:quadnorm} was calculated for all times, which includes both the kinetic energy and the potential energy \cite{radko2019thermohaline}: 
\begin{subequations}
    \label{eq:quadnorm}
\begin{align}
    e(t)&=\frac{|\hat{u}|^2 + |\hat{v}|^2 + |\hat{w}|^2}{4} + \frac{Pr|\hat{S}-\hat{T}|^2}{4 ({R}_\rho - 1)}\\
    &=:\boldsymbol{\psi}^*\boldsymbol{M}^*\boldsymbol{M}\boldsymbol{\psi}, 
\end{align}
\end{subequations}
where 
\begin{align}
\boldsymbol{M}=\text{blkdiag}\left(\sqrt{\frac{\Pr}{4(R_\rho-1)}}\begin{bmatrix}-1 & 1\end{bmatrix}, \frac{1}{2},\frac{1}{2},\frac{1}{2}\right). 
\end{align}
Here, $\text{blkdiag}(\cdot)$ means the block diagonal. This energy $e(t)$ can be connected to the vector norm $\|\boldsymbol{\psi}\|_2$ based on the Rayleigh-Ritz theorem \cite{horn2012matrix}:
\begin{align}
     e(t)\leq \mu_{\text{max}}[\boldsymbol{M}^*\boldsymbol{M}] \|\boldsymbol{\psi}(t)\|_2^2. 
\end{align}

\begin{figure}
    \centering
    \includegraphics[width=0.45\textwidth]{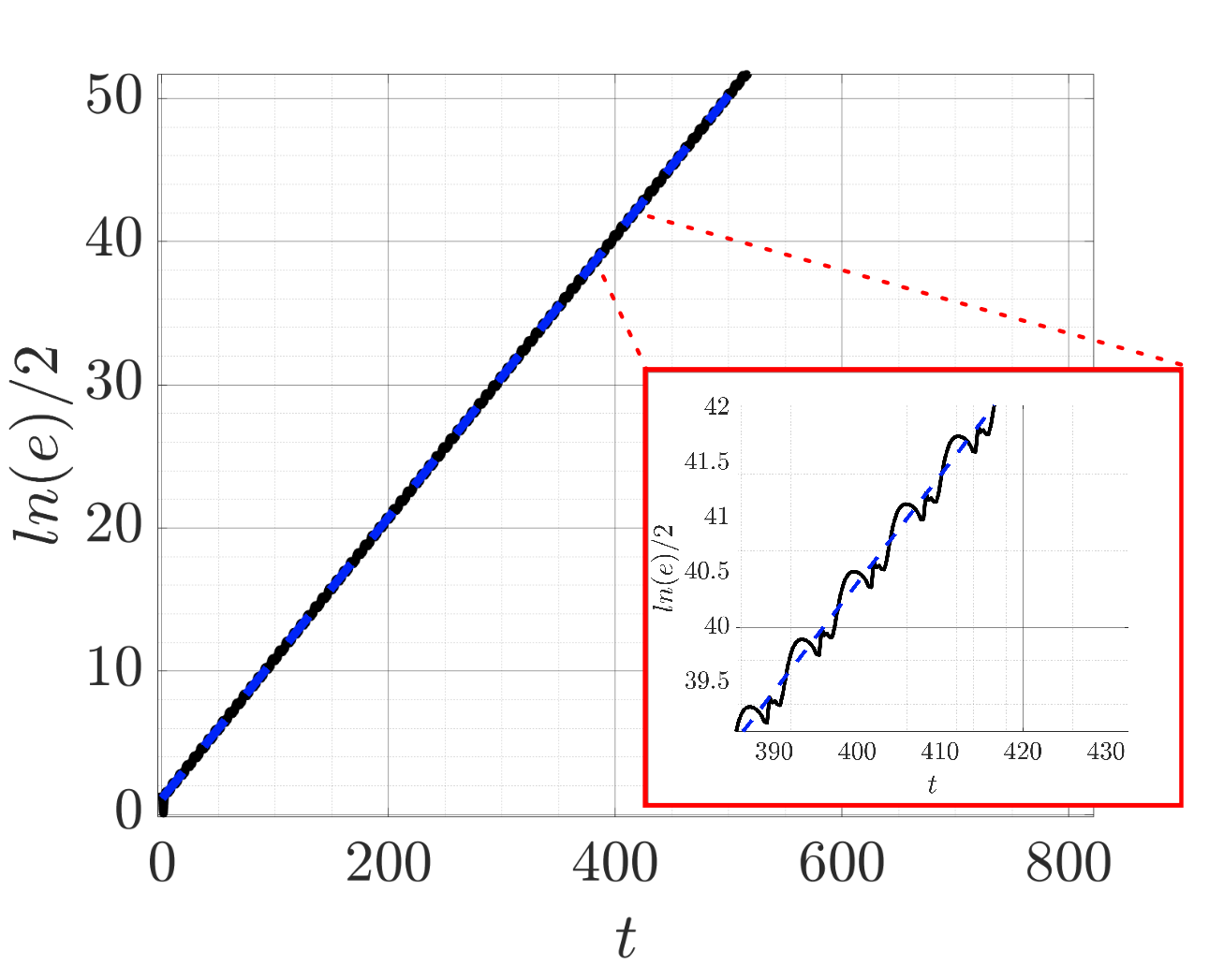}
       \caption{$\ln(e)/2$ over $t$ from numerical simulations ($\mline\mline$) of system in \eqref{eq:ode} and a least squares fitted $\ln(e)/2=\underline{\lambda}t+B$ ({\color{blue}$\dashed$}). The results are associated with wavenumbers $(k, m_0)=(0.179, 3.01 \times 10^{-3})$ and the initial condition producing the largest growth rate among 50 simulations with random initial conditions. 
       }
    \label{fig:ode45}
    \vspace{-3ex}
\end{figure}

To obtain the growth rate $\underline{\lambda}$ from numerical simulations, we then perform a least squares fit of $\ln (e)/2=\underline{\lambda} t+B$ for the second half of the time series. We only fit the second half of the time series to avoid fitting the initial transient response. \rev{For unstable regimes, which we focus on, the unstable growing modes will dominate the system response after the initial transient, while transient growth will be important for stable regimes \cite{wei2025upper}.} For each wavenumber pair, we conduct 50 simulations with initial conditions for each state variable $[\hat{T}, \hat{S}, \hat{u}, \hat{v}, \hat{w}]$ as a complex normally distributed random variable generated by \texttt{randn} command in MATLAB. We then compute the maximal growth rate from these 50 simulations. Fig. \ref{fig:ode45} shows sample results of $e(t)$ and the least squares fit associated with wavenumbers $(k, m_0)=(0.179, 3.01 \times 10^{-3})$, where this wavenumber pair is associated with the maximal growth rate \cite{radko2019thermohaline}.

\subsection{Floquet Theory}

\label{subsec:floquet}
For a linear time-varying system in Eq. \eqref{eq:ode}, we have the solution as 
\begin{equation}
\boldsymbol{\psi}(t) = \boldsymbol{\Phi}(t,t_0)\boldsymbol{\psi}(t_0),
\end{equation}
where $\boldsymbol{\Phi}(t,t_0)$ is the state-transition matrix that satisfies
\begin{equation}
\partial_t \boldsymbol{\Phi}(t,t_0) = \boldsymbol{A}(t)\boldsymbol{\Phi}(t,t_0), \quad \boldsymbol{\Phi}(t_0,t_0) = \boldsymbol{I}
\end{equation}
with $\boldsymbol{I}$ as the identity matrix. We can numerically approximate 
$\boldsymbol{\Phi}(t,t_0)$ at $t =  n\Delta t+t_0$ as
\begin{align}
&\boldsymbol{\Phi}(n\Delta t + t_0, t_0) = \prod_{j=0}^{n-1} e^{\boldsymbol{A}(j\Delta t + t_0)\Delta t}\\
=&e^{\boldsymbol{A}((n-1)\Delta t+t_0) \Delta t}\cdots e^{\boldsymbol{A}(\Delta t+t_0) \Delta t}e^{\boldsymbol{A}(t_0) \Delta t}
\end{align}
Then, we compute Floquet multiplier $\gamma:=\text{eig}[\boldsymbol{\Phi}(T+t_0, t_0)]$, where $\text{max}\,|\gamma|$ represents the growth of solution norm over one period, i.e., $\frac{\|\boldsymbol{\psi}(T+t_0)\|}{\|\boldsymbol{\psi}(t_0)\|}$. Thus, we have the temporal growth rate predicted from the Floquet theory as $\lambda_F=\ln(\text{max}\,|\gamma|)/T$.

\section{Results}
\label{sec:results}

\subsubsection{Lyapunov Method Accurately Predicts the Growth Rate} 
\begin{figure*}[!htbp]
    \centering

   \includegraphics[width=0.99\textwidth]{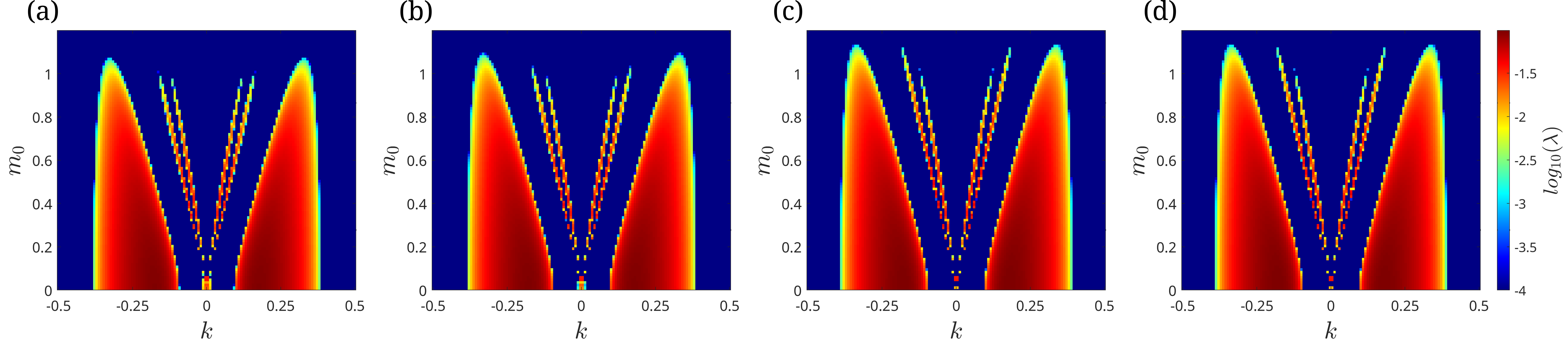}
    \caption{Stability analysis of the linear time-varying system in \eqref{eq:ode} over different $(k, m_0)$ using three different methods in \S \ref{sec:methods}. (a) $\text{log}_{10}(\overline{\lambda})$ using the Lyapunov method in \S \ref{subsec:lyapunov} with $n = 400$, (b) $\text{log}_{10}(\overline{\lambda})$ using the Lyapunov method in \S \ref{subsec:lyapunov} with $n=800$, (c) $\text{log}_{10}(\underline{\lambda})$ using numerical simulations as described in Section \ref{subsec:dns}, and (d) $\text{log}_{10}(\lambda_F)$ using the Floquet theory in \S \ref{subsec:floquet} with $n=2000$.}
    \label{fig:km0_plots}
    \vspace{-2ex}
\end{figure*}

The growth rate of the linear time-varying system in \eqref{eq:ode} was computed over initial vertical wavenumbers $m_{0} \in[0, 1.5]$ and $x$-direction wavenumbers $k \in[-0.5, 0.5]$ with uniformly sampled points $n_{\text{sample}}=126$ \rev{for} each wavenumber domain. Figs. \ref{fig:km0_plots}a and \ref{fig:km0_plots}b show $\overline{\lambda}$ obtained from the Lyapunov method in \S \ref{subsec:lyapunov} with temporal discretization points $n=400$ and $n=800$, respectively. For comparison, Fig. \ref{fig:km0_plots}c shows $\underline{\lambda}$ obtained from numerical simulations in \S \ref{subsec:dns}, and Fig. \ref{fig:km0_plots}d shows $\lambda_F$ obtained by the Floquet method in \S \ref{subsec:floquet}. In Fig. \ref{fig:km0_plots}, the growth rate results with $\log_{10}(\lambda)< -4$ were truncated to $\log_{10}(\lambda)=-4$, which are shown as dark blue.

The general structure of the unstable parameter regime (``bulbs") in each panel of Fig. \ref{fig:km0_plots} is consistent across these three methods and also consistent with simulation results in Ref. \cite{radko2019thermohaline}. There is a symmetry across the $k=0$ axis with one main bulb followed by several thinner, elongated, and disconnected parameter regimes of instability. Small initial vertical wavenumbers $m_0$ generally resulted in the greatest instability. The width and height of the unstable parameter regime in Fig. \ref{fig:km0_plots}a ($n=400$) are slightly lower than those shown in Fig. \ref{fig:km0_plots}b ($n=800$). This indicates that the Lyapunov method requires large enough temporal discretization points $n$ to predict the growth rate accurately. With increasing $n$, the Lyapunov method approached the results of numerical simulations (Fig. \ref{fig:km0_plots}c) and the Floquet theory (Fig. \ref{fig:km0_plots}d). We have also observed that the Lyapunov method with drastic under-sampling on the order of $n\sim \mathcal{O}(10^{1})$ can lose entirely the instability pattern of $\overline{\lambda}(k,m_0)$.     

Table \ref{table:maxgrowth2} shows the maximal growth rate over $(k,m_0)$ parameter regimes in Fig. \ref{fig:km0_plots} and the corresponding $(k,m_0)$ wavenumber pair for these three methods. The maximal growth rate predicted from the Lyapunov method is approaching that obtained from the numerical simulations ($\text{log}_{10}[\underline{\lambda}]=-1.00754$) and the Floquet theory ($\text{log}_{10}[\lambda_F]=-1.00765$). Moreover, the wavenumber pair associated with the maximal growth rate is $(k,m_0)=(0.180,0.000)$ for the Lyapunov method with $n=800$, which agrees well with that obtained from numerical simulations and the Floquet theory. 

\begin{table}[!htbp]
\begin{tabularx}{\columnwidth}{|l|c|X|X|}
\hline
\textbf{Method} & \textbf{${n}$} & ${\log_{10}(\lambda_\text{max})}$ & ${(k, m_0)}$ pair \\ \hline

\multirow{2}{*}{Lyapunov} 
    & 400 & -1.01807 & (0.172, 0.000) \\ \cline{2-4}
    & 800 & -1.01316 & (0.180, 0.000) \\ \hline
Simulations   & --   & -1.00754  & (0.180, 0.000) \\ \hline
Floquet & 2000 & -1.00765 
& (0.180, 0.000) \\ \hline
\end{tabularx}
\caption{Maximum growth rate and the associated wavenumber pair $(k, m_0)$ obtained by three different stability analysis methods. }
\label{table:maxgrowth2}
\vspace{-1ex}
\end{table}

\begin{figure}[!htbp]
    \centering
    \includegraphics[width=0.9\columnwidth]{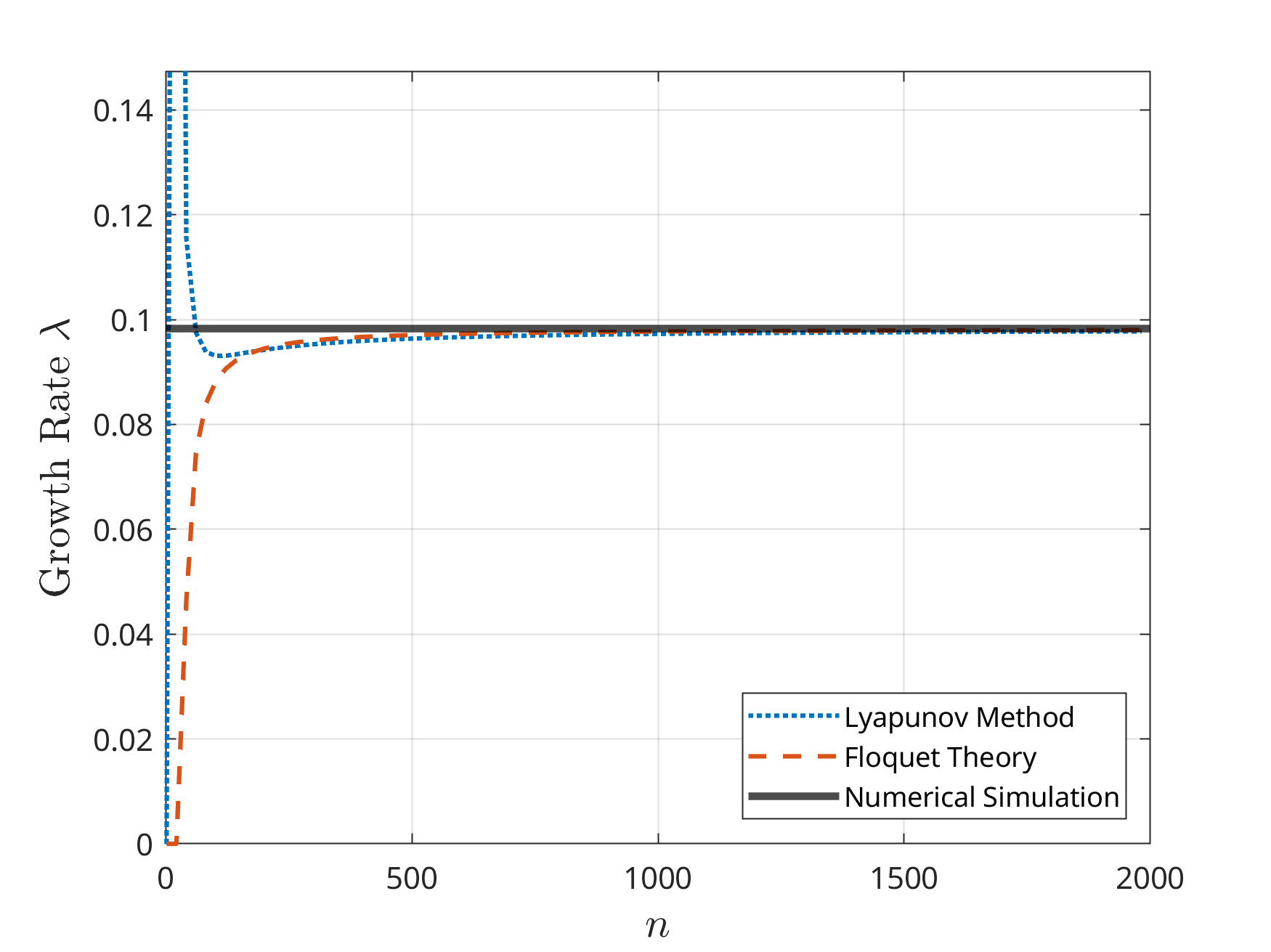}
    \caption{The convergence of the Lyapunov method and the Floquet theory for $(k, m_0)=(0.179, 3 \times 10^{-3})$ as increasing $n$, the number of the temporal discretization points. The thick black line shows the growth rate obtained by numerical simulations in \S\ref{subsec:dns}.}
    \label{fig:convergence}
    \vspace{-3ex}
\end{figure}

    Fig. \ref{fig:convergence} then investigates the convergence of the Lyapunov method and Floquet theory by plotting the growth rate as a function of $n$. We use the growth rate computed from numerical simulations ($\mline\mline$) as a basis for comparison. As the number of temporal discretization points increases, Fig. \ref{fig:convergence} shows that the growth rate predicted by the Lyapunov method $({\color{cyan}\dotted})$ and the Floquet theory $({\color{red}\dashed})$ both converge to the growth rate obtained by numerical simulations ($\mline\mline$), while Floquet theory converges faster than the Lyapunov method. 
    \rev{This suggests that undersampling of $\boldsymbol{A}(t)$ can fail to capture the essential system dynamics, causing both the Lyapunov method and Floquet theory to inaccurately estimate the growth rate. Moreover, we observe that using a Lyapunov function candidate $V=\boldsymbol{\psi}^*\boldsymbol{P}\boldsymbol{\psi}$ with a constant $\boldsymbol{P}$ cannot correctly evaluate stability for time-varying systems, indicating an important role of the time dependence of $\boldsymbol{P}(t)$ term and the $\dot{\boldsymbol{P}}(t)$ term in Theorem \ref{thm:lmi_linear} for stability analysis of time-varying systems.}

\subsubsection{Eigendecomposition of $\boldsymbol{P}(t)$}

\begin{figure}[!htbp]
    \centering

   \includegraphics[width=\columnwidth]{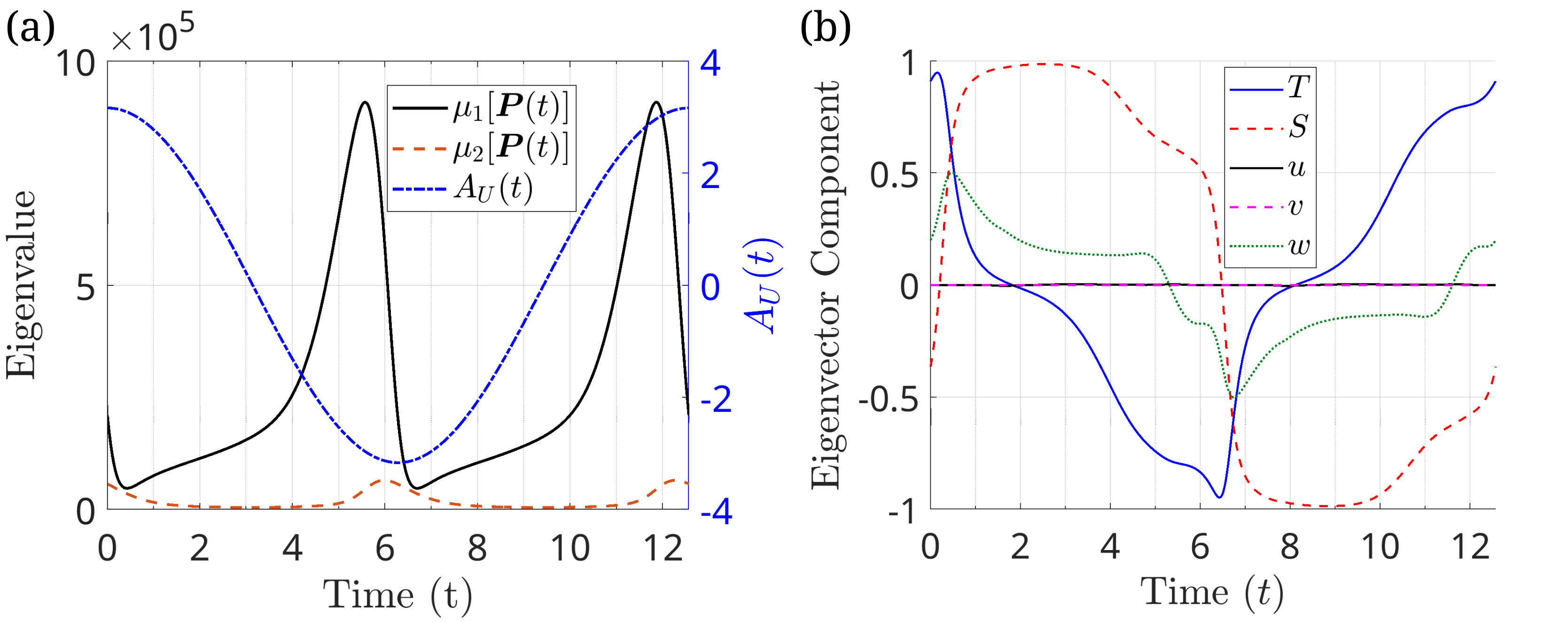}
\caption{Eigendecomposition of $\boldsymbol{P}(t)$ obtained from the Lyapunov method for $(k, m_{0})=(0.179, 3\times 10^{-3})$ with $n=2000$. (a) The two largest eigenvalues $\mu_1[\boldsymbol{P}(t)]$ and $\mu_2[\boldsymbol{P}(t)]$ over time with the base flow $A_U(t)$. (b) Eigenvector components corresponding to the $\mu_1[\boldsymbol{P}(t)]$ over time. Note that $u$ and $v$ components are nearly zero.} 
    \label{fig:eigen_analysis}
    \vspace{-3ex}
\end{figure}     

      Here, we perform an eigendecomposition of $\boldsymbol{P}(t)$ obtained from the Lyapunov method to uncover the most dangerous disturbance. \rev{We compute eigenvalues and eigenvectors of $\boldsymbol{P}(t)$ at each time $\boldsymbol{P}(t)\boldsymbol{q}_i(t)=\mu_i[\boldsymbol{P}(t)]\,\boldsymbol{q}_i(t)$, $i=1,2,\cdots 5$, leading to the eigendecomposition $\boldsymbol{P}(t)= \boldsymbol{Q}(t) \boldsymbol{\Lambda}(t) \boldsymbol{Q}(t)^{*}$ with $\boldsymbol{\Lambda}(t)=\text{diag}(\mu_1,\mu_2,\mu_3,\mu_4, \mu_5)$ and a unitary matrix $\boldsymbol{Q}(t)=[\boldsymbol{q}_1,\boldsymbol{q}_2,\boldsymbol{q}_3,\boldsymbol{q}_4,\boldsymbol{q}_5]$. Eigendecomposition of $\boldsymbol{P}(t)$ at} each time $t$ was performed using the \texttt{eigenshuffle} routine \cite{derrico_eigenshuffle_2025} to ensure continuity and consistency of eigenvectors and eigenvalues over time.  \rev{This eigendecomposition allows us to write ${V}(\boldsymbol{\psi},t)=\boldsymbol{\psi}^*\boldsymbol{P}(t)\boldsymbol{\psi}= \boldsymbol{\psi}^*\boldsymbol{Q}(t) \boldsymbol{\Lambda}(t) \boldsymbol{Q}(t)^{*}\boldsymbol{\psi}=\sum_{i=1}^5\mu_i[\boldsymbol{P}(t)]|\alpha_i|^2$, where $\boldsymbol{\psi}=\sum_{i=1}^5\alpha_i \boldsymbol{q}_i=\boldsymbol{Q}\boldsymbol{\alpha}$ with $\boldsymbol{\alpha}:=[\alpha_1,\cdots,\alpha_5]^{\text{T}}$. Based on the proof of Theorem \ref{thm:lmi_linear}, we have $V(\boldsymbol{\psi}(t),t)\leq e^{2\overline{\lambda} (t-t_0)}V(\boldsymbol{\psi}(t_0),t_0)$, and thus, it is most dangerous to reach the upper bound of $V(\boldsymbol{\psi}(t),t)$ when disturbing the system at the moment $t_m$ corresponding to the peak of the largest eigenvalue $\mu_1[\boldsymbol{P}(t)]$ using the associated eigenvector $\boldsymbol{q}_1(t_m)$. As a result, we interpret eigenvector $\boldsymbol{q}_1(t_m)$ as the most dangerous disturbance.}

      Fig. \ref{fig:eigen_analysis}a shows the largest two eigenvalues $\mu_1$ and $\mu_2$ of $\boldsymbol{P}(t)$ \rev{corresponding to $n=2000$ and the wavenumber pair $(k, m_{0})=(0.179, 3\times 10^{-3})$}. The peak $\mu_1[\boldsymbol{P}(t)]$ aligns closely with the peak strength of the background shear flow $A_U(t)$, \rev{suggesting that disturbing the flow at this moment will be most dangerous to reach the upper bound of $V$.} Fig. \ref{fig:eigen_analysis}b then shows each component of the eigenvector $\boldsymbol{q}_1(t)$ associated with $\mu_1[\boldsymbol{P}(t)]$. The $v$ velocity component is zero for all time since a 2D model was studied (i.e., $l=A_V(t)=0$), and the $u$ velocity component was also negligible. \rev{At the moment passing through the peak of $\mu_1[\boldsymbol{P}(t)]$ ($t\in [4,6.5]$), the magnitude of the temperature mode $T$ increases, while the salinity mode $S$ decreases slowly and then sharply. This suggests that the most dangerous disturbance is dominated by the temperature mode, which is consistent with the physics that the background temperature $\overline{T}=-z$ is destabilizing. }

\subsubsection{Computational Resources Requirement of the Lyapunov Method}
    
    We also test the memory usage and running time for the Lyapunov method as summarized in Table \ref{table:perf}. Here, we parallelize the computation in the $k$ parameter regime and use one computing node with 126 cores and 500~GB of RAM at UConn Storrs High Performance Computing (HPC) facility to generate results in Fig. \ref{fig:km0_plots}. We take $n_{\text{sample}}=126$ samples in both $k$ and $m_0$ parameter regimes. As the number of temporal discretization points $n$ increases, the computational time and memory requirements of the Lyapunov method increase as a result of the increased number of constraints. The CPU hours increase much faster than the memory usage over the number of temporal discretization points $n$. Note that the time per $(k, m_0)$ and the memory per CPU are estimates of the average usage across all CPUs, whether or not they were idle. Because we do not solve the growth rate $\overline{\lambda}$ for the stable case if the constraints in \eqref{eq:lmi_all_discretized} are feasible with $\overline{\lambda}=0$ (i.e., the system is stable), idle CPUs could have occurred if their allotted work contains more stable parameter regimes than other CPUs. We also report the computational resources required for numerical simulations and the Floquet theory in Tab. \ref{table:perf}. Numerical simulations (with 50 random initial conditions) require computational time similar to the Lyapunov method with $n\in [200,400]$, while the Floquet theory will be much faster than the Lyapunov method.

\begin{table}[!htbp]
\renewcommand{\arraystretch}{1.3} 
\begin{tabular}{|>{\centering\arraybackslash}m{1.2cm}|
                    >{\centering\arraybackslash}m{0.5cm}|
                    >{\centering\arraybackslash}m{1.35cm}|
                    >{\centering\arraybackslash}m{1.35cm}|
                    >{\centering\arraybackslash}m{0.75cm}|
                    >{\centering\arraybackslash}m{0.8cm}|}
\hline
\textbf{Method} & \textbf{$n$} &
\textbf{Max Mem. Usage (GB)} &
\textbf{Mem. per CPU (GB)} &
\textbf{CPU Hours} &
\textbf{Time per $(k, m_0)$ (s)} \\ \hline
\multirow{4}{*}{Lyapunov} 
 & 200 & 244.5 & 1.94  & 235  & 53.2  \\ \cline{2-6}
 & 400 & 332.4 & 2.63 & 693 & 157 \\ \cline{2-6}
 & 600 & 328.1  &  2.60 & 1638 & 371.5 \\ \cline{2-6}
 & 800 & 354.3 & 2.81 &  2685 & 690  \\ \hline
Simulations   & --   & 196.1 & 1.56 & 464.7 & 105  \\ \hline
Floquet & 2000 & 187.4 & 1.49 & 5.04  & 1.14 \\ \hline
\end{tabular}
\caption{HPC performance for each method corresponding to Fig.~\ref{fig:km0_plots}. The maximal memory usage and CPU hours are for the computation of $n_{\text{sample}}=126$ horizontal wavenumber $k$ and $n_{\text{sample}}=126$ initial vertical wavenumber $m_0$. We parallelize the code in the $k$ parameter regime with 126 cores. Performance of numerical simulation is associated with 50 random initial conditions.}
\label{table:perf}
\vspace{-1ex}
\end{table}

Although Floquet theory is much more computationally efficient, it is limited to analyzing linear time-periodic systems. The Lyapunov method, instead, can be used to analyze the stability of nonlinear systems \cite{liu2020input} and non-periodically time-varying systems \cite{wei2025upper}. This work serves as a first step to demonstrate that the Lyapunov approach can predict the growth rate consistent with numerical simulations and Floquet theory for linear time-periodic systems. 
      
\section{Conclusion and Future Work}
\label{sec:conclusion}
    We demonstrated that the Lyapunov method is a viable tool to identify instabilities for a linear time-periodic fluid system describing thermohaline convection with a periodic time-varying background shear flow. In the linear matrix inequalities formulation, we include the time derivative of the weighting matrix $\dot{\boldsymbol{P}}(t)$ of the Lyapunov function candidate, which is approximated by the forward Euler method. With enough temporal discretization points, we find that the growth rate predicted by the Lyapunov method or the Floquet theory converges to that obtained by numerical simulations. The eigenvectors of the weighting matrix $\boldsymbol{P}(t)$ are employed to uncover the most dangerous disturbances. 

    We also benchmark the memory and computational time required by the Lyapunov method and how it varies over the number of temporal discretization points, which suggests future directions for improving the computational efficiency of the Lyapunov method. \rev{Another future direction is to investigate other high-order numerical methods to approximate $\dot{\boldsymbol{P}}(t)$ in LMI. Moreover, we also aim to leverage this work as a building block to extend the Lyapunov method to analyze the nonlinear stability \cite{liu2020input}, nonlinear input-output analysis \cite{wei2025nonlinear}, and non-periodically time-varying systems \cite{wei2025upper}.}
    
\bibliographystyle{IEEEtran}
\bibliography{refs}

@article{kochnev2025thesis,
  title={Stability Analysis of Thermohaline Convection With a Time-Varying Shear Flow Using the {L}yapunov Method},
  author={Kochnev, Kalin},
  journal={Honors Scholar Theses},
  number={1149},
  year={2026},
  url={https://digitalcommons.lib.uconn.edu/srhonors_theses/1149/}
}

@article{wei2025nonlinear,
  title={Nonlinear input-output analysis of transitional shear flows using small-signal finite-gain $\mathcal{L}_p$ stability},
  author={Wei, Zhengyang and Liu, Chang},
  journal={Phys. Rev. Fluids},
  volume={10},
  number={10},
  pages={103903},
  year={2025},
  publisher={APS}
}

@book{horn2012matrix,
  title={Matrix analysis},
  author={Horn, Roger A and Johnson, Charles R},
  year={2012},
  publisher={Cambridge university press}
}

@inproceedings{Lofberg2004,
  author = {Lofberg, J.},
  title = {{YALMIP}: A toolbox for modeling and optimization in {MATLAB}},
  booktitle = {Proceedings of the IEEE International Conference on Robotics and Automation},
  volume = {3},
  pages = {284--289},
  year = {2004}
}

@manual{Mosek2024,
  author = {{MOSEK ApS}},
  title = {The MOSEK optimization toolbox for MATLAB manual},
  year = {2024}
}

@article{zhou2011periodic,
  title={Periodic {L}yapunov equation based approaches to the stabilization of continuous-time periodic linear systems},
  author={Zhou, Bin and Duan, Guang-Ren},
  journal={IEEE Trans. Autom. Control	},
  volume={57},
  pages={2139--2146},
  year={2011},
  publisher={IEEE}
}

@article{bohm2012stability,
  title={Stability of periodically time-varying systems: Periodic {L}yapunov functions},
  author={B{\"o}hm, Christoph and Lazar, Mircea and Allg{\"o}wer, Frank},
  journal={Automatica},
  volume={48},
  pages={2663--2669},
  year={2012},
  publisher={Elsevier}
}

@article{graham2021exact,
  title={Exact coherent states and the nonlinear dynamics of wall-bounded turbulent flows},
  author={Graham, Michael D and Floryan, Daniel},
  journal={Annu. Rev. Fluid Mech.},
  volume={53},
  pages={227--253},
  year={2021},
  publisher={Annual Reviews}
}

@article{knobloch2014stability,
  title={Stability on time-dependent domains},
  author={Knobloch, Edgar and Krechetnikov, Rouslan},
  journal={J. Nonlinear Sci.},
  volume={24},
  pages={493--523},
  year={2014},
  publisher={Springer}
}

@article{alford2016near,
  title={Near-inertial internal gravity waves in the ocean},
  author={Alford, Matthew H and MacKinnon, Jennifer A and Simmons, Harper L and Nash, Jonathan D},
  journal={Annu. Rev. Mar. Science},
  volume={8},
  pages={95--123},
  year={2016},
  publisher={Annual Reviews}
}

@book{sutherland2010internal,
  title={Internal gravity waves},
  author={Sutherland, Bruce R},
  year={2010},
  publisher={Cambridge university press}
}

@article{liu2020input,
  title={Input-output inspired method for permissible perturbation amplitude of transitional wall-bounded shear flows},
  author={Liu, Chang and Gayme, Dennice F},
  journal={Phys. Rev. E},
  volume={102},
  pages={063108},
  year={2020},
  publisher={APS}
}

@article{joglekar2015geometry,
  title={Geometry of the edge of chaos in a low-dimensional turbulent shear flow model},
  author={Joglekar, Madhura and Feudel, Ulrike and Yorke, James A},
  journal={Phys. Rev. E},
  volume={91},
  pages={052903},
  year={2015},
  publisher={APS}
}

@article{washam2020tidal,
  title={Tidal modulation of buoyant flow and basal melt beneath {P}etermann {G}letscher {I}ce {S}helf, {G}reenland},
  author={Washam, Peter and Nicholls, Keith W and M{\"u}nchow, Andreas and Padman, Laurie},
  journal={J. Geophys. Res.: Oceans},
  volume={125},
  pages={e2020JC016427},
  year={2020},
  publisher={Wiley Online Library}
}

@article{wei2025upper,
  title={Upper bound of transient growth in accelerating and decelerating wall-driven flows using the {L}yapunov method},
  author={Wei, Zhengyang and Zhao, Weichen and Liu, Chang},
  journal={arXiv preprint arXiv:2508.01410},
  year={2025}
}

@article{davis1976stability,
  title={The stability of time-periodic flows},
  author={Davis, Stephen H},
  journal={Annu. Rev. Fluid Mech.},
  volume={8},
  pages={57--74},
  year={1976},
  publisher={Annual Reviews 4139 El Camino Way, PO Box 10139, Palo Alto, CA 94303-0139, USA}
}

@article{radko2019thermohaline,
  title={Thermohaline-shear instability},
  author={Radko, Timour},
  journal={Geophys. Res. Lett.	},
  volume={46},
  pages={822--832},
  year={2019},
  publisher={Wiley Online Library}
}

@article{radko2019instabilities,
  title={Instabilities of a time-dependent shear flow},
  author={Radko, Timour},
  journal={J. Phys. Oceanogr.},
  volume={49},
  pages={2377--2392},
  year={2019}
}

@book{Boyd1994,
  author = {Boyd, S. and El Ghaoui, L. and Feron, E. and Balakrishnan, V.},
  title = {Linear Matrix Inequalities in System and Control Theory},
  publisher = {Society for Industrial and Applied Mathematics},
  year = {1994}
}

@book{khalil2002nonlinear,
  title={{Nonlinear systems}},
  author={Khalil, H. K.},
  publisher={Upper Saddle River},
  year={2002}
}

@inproceedings{floquet1883equations,
  title={{Sur les {\'e}quations diff{\'e}rentielles lin{\'e}aires {\`a} coefficients p{\'e}riodiques}},
  author={Floquet, Gaston},
  booktitle={Ann. Sci. Ec. Norm. Super.},
  volume={12},
  pages={47--88},
  year={1883}
}

@online{derrico_eigenshuffle_2025,
	title = {Eigenshuffle},
	url = {https://www.mathworks.com/matlabcentral/fileexchange/22885-eigenshuffle},
	titleaddon = {Eigenshuffle - {MATLAB} Central File Exchange},
	author = {D'Errico, John},
	urldate = {2025-09-14},
	date = {2025-09-14},
	langid = {english},
}

\end{document}